\documentclass[prl,twocolumn,aps,showpacs]{revtex4}
\usepackage{graphicx}
\begin{document}

\title{Ultracold atoms in optical lattices with random on-site interactions}

\author{H. Gimperlein$^1$, S. Wessel$^1$, J. Schmiedmayer$^2$, and 
L. Santos$^1$}
\affiliation{
(1) Institut f\"ur Theoretische Physik III, 
Universit\"at Stuttgart, Pfaffenwaldring 57 V, D-70550 Stuttgart \\
(2) Physikalisches Institut, Universit\"at Heidelberg, 
Philosophenweg 12, D-69120 Heidelberg
} 
\begin{abstract}

We consider the physics of lattice bosons affected by disordered on-site interparticle interactions.
Characteristic qualitative changes in the zero temperature phase diagram are observed   
when compared to the case of randomness in the chemical potential. 
The Mott-insulating regions shrink and  eventually vanish
for any finite disorder strength beyond a sufficiently large filling factor. 
Furthermore, at low values of the chemical potential both the superfluid and Mott insulator 
are stable towards formation of a Bose glass
leading to a possibly non-trivial tricritical point. 
We discuss feasible experimental realizations of our scenario 
in the context of ultracold atoms on optical lattices.
\end{abstract}

\maketitle


Ultracold atomic gases have attracted large interest during the last
years, in particular due to the experimental achievement of 
quantum degeneracy both for bosonic \cite{BEC} and fermionic \cite{Fermi} 
gases. Very recently, the extraordinary experimental developments in 
the manipulation of ultracold atoms have opened the path towards a 
new fascinating research area, namely the analysis of strongly-correlated 
atomic gases. In this sense, remarkable experimental results have 
been already reported, including experiments on the so-called 
BEC-BCS crossover in degenerate Fermi gases \cite{BCSBEC}, and 
experiments on ultracold gases in optical lattices, such as e.g. 
the observation of the Mott insulator-to-superfluid transition \cite{Greiner,ETH} and the 
experimental realization of a Tonks-Girardeau gas \cite{Paredes}.

Up to now the experiments performed in optical lattices have considered 
a defect-free square lattice potential. However, recent proposals have 
discussed new scenarios which depart from this standard. In particular, 
various lattice geometries are achievable 
by standard laser techniques, and should lead to fascinating physics~\cite{Kagome,triangular}. 
On the other hand, disorder may be induced in the laser potential in 
different ways. Localized impurities can be created, leading 
to Kondo-like physics~\cite{Cirac}. Randomness 
can be produced by means of additional incommensurable lattices, 
or by laser speckles~\cite{BG}, and may lead to 
Anderson localization~\cite{Anderson} and Bose glass phases~\cite{Fisher}.
First experimental steps towards random potentials 
have already been achieved~\cite{Inguscio}. 
Additionally, randomness may be induced in the hopping 
rates in the optical lattice, leading to Fermi glasses, spin glasses 
and quantum percolation~\cite{Sanpera}. 

One of the most fascinating possibilities for the control of 
atomic gases is provided by the manipulation of the interatomic interactions 
by means of Feshbach resonances~\cite{Feshbach}. Using a suitably adjusted magnetic field a resonance 
is induced between a molecular state and the unbound states of the incoming atoms, leading 
to important modifications of the scattering properties. 
The $s$-wave scattering length undergoes a very large change in a relatively 
narrow window of values of the applied magnetic field, becoming 
$\pm\infty$ at the resonance. This extreme sensitivity of the scattering 
properties at the verge of a Feshbach resonance is crucial for our
discussion.

In this Letter, we consider the physics of lattice bosons subject to 
a novel kind of disorder, namely bounded disorder in 
the strength of the interatomic interactions~\cite{Adrian}. 
In particular, we show how this type of disordered system can be realized in the
context of ultracold atoms near a Feshbach resonance. 
We discuss the consequences that the disorder in the interaction strength 
has on the zero-temperature phase diagram of the system 
by means of a strong-coupling expansion (SCE) and quantum Monte Carlo (QMC) simulations, and 
contrast our findings to 
the case of randomness in the chemical potential~\cite{Fisher}.

\begin{figure}
\includegraphics[width=5.8cm]{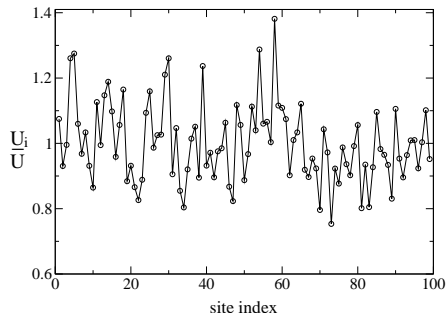}
\caption{Typical spatial variation of the on-site interactions.
In the figure we have considered a $^{87}$Rb cloud  
$0.5\mu$m over the surface of an atom chip, at a $1\mu$m-wide magnetic 
wire with a current leading to a magnetic field 
variation of up to $\pm 2.5$mG, an off-set field $5mG$ away from the 
$685.8$G resonance, 
a lattice spacing of $1\mu$, and a lattice depth of $5$ recoil energies. 
$U_i$ is obtained by integrating
$\int dx U(x)|\psi_i(x)|^4/\int dx |\psi_i(x)|^4$, with $\psi_i(x)$ the $i$-th on-site 
wavefunction which is approximated by a Gaussian.}
\label{fig:fig1}
\end{figure}

The above scenario can be realized in 
a gas of ultracold bosons confined to 
an optical lattice using state-of-the-art experimental techniques on atom chips \cite{Schmiedmayer}. 
In the following, we consider a one-dimensional 
configuration, where the atoms are assumed to be strongly confined in the other 
two dimensions (by e.g. additional optical confinement). 
However, qualitatively similar results as those discussed in this Letter 
also hold in higher dimensions. We assume 
that the gas is brought at the verge of a Feshbach resonance by an off-set magnetic field,
where, as  discussed previously, slight modifications of
the magnetic field lead to large variations of the scattering properties.
We furthermore consider the  bosonic gas to be close to a magnetic wire, inducing 
a spatially random magnetic field \cite{Schmiedmayer}, which can be considered 
sufficiently weak, such that the variations in the potential energy can be considered 
negligible when compared to other energy scales. However, since the off-set 
field sets the system at the verge of a Feshbach resonance, the 
slight random variations of the magnetic field lead to a spatially 
random variation of the local interatomic interactions (see Fig.~\ref{fig:fig1}). 
We assume that the variations of the scattering properties are bounded, i.e. the spatial
variations of the magnetic field do not induce a crossing of the resonance. 

In order to avoid significant van der Waals losses, the atoms
should be placed at a distance $>0.5\mu$m from the surface of the atom chip,
which restricts the characteristic wavelength of the variations of the magnetic
field at the wire to $\Delta x \ge 1\mu$m. The inter-site separation in
the optical lattice is approximately $\lambda=400$nm, but can be
increased to $1\mu$m by setting an angle between the counter-propagating lasers in
such a way that $\Delta x \le \lambda$. The variation of the magnetic field at the
wire can be adapted to typical Feshbach resonance widths, which 
are of the order of $1-10$ mG. Fig.~\ref{fig:fig1} shows an example of the  
variation of the on-site interaction for a typical experimental setup.

Under these conditions, the system is described by a Bose-Hubbard Hamiltonian 
with random interatomic interactions:
\begin{equation}
\hat H= -J\sum_{<ij>}\left ( b_i^\dag b_j +h.c.\right )+
\sum_i \frac{U_i}{2}n_i(n_i-1)-\sum_i \mu_i n_i,
\end{equation}
where $b_i$ ($b_i^\dag$) is the annihilation (creation) operator 
for bosons in the $i$-th lattice site, $n_i=b_i^\dag b_i$, $J$ is the 
hopping constant, $\mu_i=\mu$ is the local chemical potential, and $U_i$ 
denotes the local value of the coupling constant for the interatomic interactions. 
We assume $U_i$ uniformly distributed inside the range $U(1-\epsilon)\le U_i \le U(1+\epsilon)$  \cite{footnote}.

In order to assess the effects of randomness in the on-site interactions and to contrast it to the case of a
random chemical potential,
we shortly review the latter case, where
$\mu-\Delta\le\mu_i\le\mu+\Delta$, at a fixed  $U_i=U$ ($\epsilon=0$)~\cite{Fisher}: 
The phase diagram in the strongly interacting regime can be obtained from considering first the
trivial limit, $J=0$:
One finds that if $\mu$ falls within an interval $(n-1)U+\Delta\leq\mu\leq nU-\Delta$, with an integer $n$,
the ground-state of the
system is an incompressible Mott insulator with $n$ bosons
on each lattice site. For values of $\mu$ outside these intervals, i.e. for 
$nU-\Delta<\mu<nU+\Delta$, 
the system
enters an insulating, but compressible phase commonly referred to as
Bose glass, in which the occupation per site varies between
$n$ and $n+1$. 
This physics is maintained up to  moderate values of $J$, leading to a phase diagram with three distinguished phases:
superfluid (SF) at large $J$, Bose glass (BG), and the Mott insulator (MI) regions. Recent
quantum Monte Carlo calculations have shown that the transition from SF to MI
always occurs through an intermediate BG phase \cite{Svistunov}, as suggested
in Ref.~\cite{Fisher}.

A characteristic difference to the case of random $U$ at a fixed value of $\mu_i=\mu$ ($\Delta=0$), 
is already obtained in the limit $J=0$: Now compressible 
phases occur for $n(1-\epsilon)<\mu/U<n(1+\epsilon)$, in which the occupation
per site varies between $n$ and $n+1$, whereas outside these regions the system enters into a MI.
In contrast to the case of randomness in the chemical potential, 
the extent of compressible regions between the MI thus increases with the occupation number $n$.
The disorder in $U$ thus leads to an increased destabilization of MI states with increasing $n$.
Eventually, MI regions with an occupation of $n\ge (1+\epsilon)/2\epsilon$
particles per site disappear. Therefore, given a disorder strength $\epsilon$, only a finite number of MI regions 
remain stable. 
This selective destruction of MI states cannot be accessed in the case of a random chemical potential, where {\it all} the MI lobes vanish
once $\Delta\geq U/2$.
More generally, if both $U_i$ and $\mu_i$ 
have a bounded random distribution, the $n$-th lobe disappears for
$(2n-1)\epsilon+2\Delta/U=1$.


In order to analyze the phase diagram for finite hopping $J>0$, we first estimate the extent of the MI regions 
using the SCE~\cite{Monien}. This method allows for a quantitative
calculation
of the boundaries between compressible and incompressible phases for sufficiently 
low $J$, particularly in low dimensions, where the coordination number is small. 


From the SCE one obtains the energy gap for
adding or removing a particle from the MI, and performs an expansion in the 
small parameters $J/U$, $\Delta/U$ and $\epsilon$. For a given value of $J$, these gaps determine 
the boundaries of the MI lobes, which are obtained in first order as
\begin{eqnarray}
\frac{\mu_{u}}{U} &=& -2 (n + 1) \frac{J}{U} - \frac{\Delta}{U} + n (1-\epsilon) \\
\frac{\mu_{l}}{U} &=& 2 n \frac{J}{U} +\frac{\Delta}{U} + (n-1) (1+\epsilon)
\end{eqnarray}
for the upper ($u$) and lower ($l$) boundaries of the MI with integer filling $n$.
In particular we find, that the lower boundary of the first MI lobe ($n=1$) is not affected 
by the disorder in $U$, in contrast to the random-$\mu$ case~\cite{Monien}.
This reflects the observation, that no compressible phase emerges below the $n=1$ MI at $J=0$ for the random-$U$ case,
in contrast to the case of random $\mu$.

From equating the two boundaries one obtains an (under-) estimate for the largest extent of the MI phase with
filling $n$:
\begin{equation}
\frac{J_c}{U}=\frac{1-2\Delta/U-(2n-1)\epsilon}{2(2n+1)}.
\end{equation} 

Calculating higher-order terms in the strong-coupling expansion improves the quantitative description of the MI lobes, 
but does not affect our qualitative conclusions. 
In particular, all the higher order terms in the random interaction are at least of order  $(J/U)^2 \epsilon$. 
Considering that the critical hopping $J_c / U \ll 1$, even relatively large disorder strengths may be approximated 
using the first order shift of the phase boundaries. 
For the figures presented below, we used expansions up to third order and also 
considered finite size effects ~\cite{Monien}. The latter, as shown below, are rather 
significant, and converge only very slowly to the thermodynamic limit when the number of lattice sites considered 
increases.

\begin{figure}
\includegraphics[width=6.0cm]{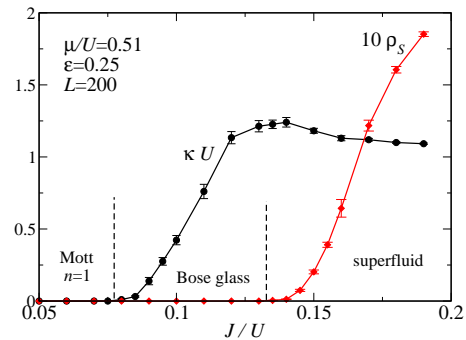}
\caption{Zero temperature compressibility $\kappa$ and superfluid density $\rho_S$ of 
bosons on a one-dimensional optical lattice with random interaction strength of $\epsilon=0.25$ 
as a function of $J/U$ along a cut a $\mu/U=0.51$, obtained from QMC simulations.
The error bars quantify the standard deviation of the sample-to-sample fluctuations, 
whereas the dashed lines indicate the estimated extent of the Bose glass
along this cut.
}
\label{fig:mucut}
\end{figure}

While the SCE allows to estimate the extent of the Mott-insulating phases quantitatively,
it does not provide information about the complete phase diagram, in particular concerning the nature of the compressible phases and the 
presence of SF and BG phases. 
In order to obtain a more complete phase 
diagram for bosons with random on-site interactions we have performed QMC simulations using the stochastic series expansion 
method~\cite{sse} with directed loop updates~\cite{directedloop1,directedloop2}. 
We performed simulations for periodic chains with $L=200$ sites
averaged over typically 200 disorder realizations for each data point. 
In the simulations, the temperature $T$
was chosen low enough to obtain ground 
state properties of the finite systems~\cite{footnoteT}. 
In order to reliably identify the various phases of the system, we measured the global compressibility 
$\kappa=\partial N / \partial \mu$
from the fluctuations of the particle number $N$, and the winding number 
fluctuations $\langle W^2\rangle$, from which the superfluid density is obtained as $\rho_S=LT/(2 J)\;\langle W^2 \rangle$~\cite{rho_S}.
Fig.~\ref{fig:mucut} 
shows the disorder averaged values of both observables 
as functions of $J/U$ along a cut of constant $\mu/U=0.51$ for $\epsilon=0.25$, which at $J=0$ belongs to the MI region with $n=1$. 
At low values of $J$, an extended MI region with $n=1$ is clearly 
identified by vanishing values of both $\kappa$ and $\rho_S$.
For large value of $J/U$, both $\kappa$ and $\rho_S$ take 
on finite values, thus identifying the large-$J$ region as a SF phase. 
In the intermediate region, for $0.078\lesssim J/U \lesssim 0.133$, the system shows a finite
compressibility, but no superfluid response. This regime is thus characterized as a disorder-induced  
BG phase, separating the 
MI and SF region at this value of $\mu/U=0.51$.

\begin{figure}
\includegraphics[width=6.0cm]{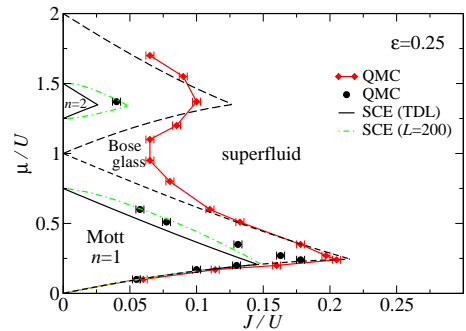}
\caption{Zero-temperature phase diagram of bosons on a one-dimensional optical lattice with random interaction strength of  $\epsilon=0.25$,
obtained from QMC simulations of $200$ sites and a third-order SCE
for the thermodynamic limit (TDL). The extent of the 
Mott lobes 
in the pure case ($\epsilon=0$) from SCE is indicated by the dashed line, whereas the dot-dashed line shows the SCE results for a finite system 
of $L=200$ sites.
}
\label{fig:phasediag}
\end{figure}

Performing similar scans along different lines in parameter space, 
we obtain QMC estimates of the various phase boundaries, leading 
to the phase diagram of Fig.~\ref{fig:phasediag}. Here, we combined the QMC results for $\epsilon=0.25$ with third-order SCE results 
for the extent of the MI phases in the thermodynamic limit (TDL). 
Within the chosen resolution, the finite size corrections of the phase boundaries obtained by QMC show only a weak dependence on the system size.
An accurate determination of the MI boundaries from QMC in the disordered case~\cite{scalettar} 
would require significantly larger system sizes due to rare regions of delocalized bosons~\cite{Monien}. In fact, performing a 
third-order SCE for a finite chain
of $L=200$ sites compares well to the estimated extend of the MI regions from QMC (Fig.~\ref{fig:phasediag}) in the strong-coupling 
regime, $J/U\lesssim 0.1$, and indicates the rather strong finite size effects in the disordered case.

Our numerical calculations confirm the qualitative picture considered above:
The MI lobes clearly shrink, not only in the $\mu$-direction but also in $J$, and eventually 
disappear for sufficiently large values of $\mu$. 
In the case considered in  Fig.~\ref{fig:phasediag}, where $\epsilon=0.25$, only the MI lobes with $n=1$ and $2$ remain stable.
Furthermore, we observe that the lower boundary of the $n=1$ MI does not vary significantly with the disorder strength
up to the critical hopping $J_c$. The reduced relevance of disorder in the interaction strength 
in the low-density region is also reflected by the complete absence of a BG phase for $\mu<0$. 
Therefore, a dilute compressible lattice boson gas
remains superfluid even in the presence of disordered interactions. 
This is in clear contrast to the case of random $\mu$,
where the BG extends down to arbitrary low but finite densities.
The absence of a BG for  $\mu<0$ indicates a tricritical point for the SF, MI, and BG phases 
along the lower boundary of the $n=1$ MI. 
However, our calculations cannot 
reliably distinguish whether the BG phase disappears exactly 
at a tricritical point at $\mu=J=0$, or if this occurs for finite values of $\mu,J>0$.
This interesting question thus remains open for future studies.


To summarize, in this Letter we analyzed the novel physics occurring when the on-site interaction
in a Bose-Hubbard Hamiltonian acquires a bounded random character. We discussed how this exciting possibility can be realized in 
ultracold atomic gases in optical lattices at the verge of a Feshbach resonance, when they are brought in the vicinity of weak 
spatially random magnetic fields generated by magnetic wires. As for the case of random chemical potentials, 
a phase diagram with superfluid, Bose-glass and Mott-insulator phases is predicted. However, important differences 
can be found when comparing the random $U$ and random $\mu$ cases: For random interactions,  the Mott-insulator 
phases become progressively narrower (both in the $\mu$ and $J$ direction) for larger occupation numbers, and 
eventually disappear beyond a given filling. Furthermore, the 
Bose-glass phase disappears for low chemical potentials and hopping, and hence a tricritical point occurs, 
although our numerical calculations cannot discern whether this occurs at $\mu=J=0$ or very 
small but finite values.

Our calculations were performed in absence of an overall (usually harmonic) 
confining potential, whereas the necessary commensurability for Mott-insulator 
phases can be achieved in practice only in the presence of inhomogeneous potentials. However, if the 
confining potential is sufficiently shallow, a local chemical potential may be considered, and local 
phase occur, as thoroughly demonstrated by means of QMC calculations~\cite{Rigol,Wessel} for the pure case. 
Of course, the finite size of these different regions could significantly affect the boundaries between the Bose glass and 
superfluid, once the localization length becomes smaller than the spatial extent of the various regions.
Typical experiments on the Mott-insulator phase of atoms in optical lattices~\cite{Greiner,ETH} rely on the 
observation of the broadening of interference fringes in time-of-flight pictures 
(provided by the insulating character of the phase), and the monitoring of the opening of the Mott-gap in the 
excitation spectrum by means of tilting experiments. In these experiments, although other spatial 
phases are expected, the Mott-insulator region is the largest one and dominates the experimental detection. 
Similar experiments could be performed to detect the shrinking of the Mott-insulator phases as a function of the 
disorder in $U$ predicted in this paper. 

Fruitful conversations with A. Kantian, M. Lewenstein, and P. \"Ohberg are 
acknowledged. This work was supported by the Alexander von Humboldt Foundation and by
NIC at FZ J\"ulich.
H.G. thanks the Studienstiftung des deutschen Volkes for support.


\begin{thebibliography}{99}

\bibitem{BEC} M. H. Anderson, {\it et al.},Science {\bf 269}, 198 (1995).

\bibitem{Fermi} B. DeMarco and D. S. Jin, Science {\bf 285}, 1703 (1999).

\bibitem{BCSBEC} C. A. Regal, M. Greiner, and D. S. Jin, Phys. Rev. Lett. {\bf 92}, 040403 (2004);
M. W. Zwierlein, {\it et al.}, Phys. Rev. Lett. {\bf 92}, 120403 (2004).

\bibitem{Greiner} M. Greiner \textit{et al.}, Nature \textbf{415}, 39 (2002).

\bibitem{ETH} T. St\"oferle, H. Moritz, G. Schori, M. K\"ohl and T. Esslinger,
Phys. Rev. Lett. {\bf 91}, 130403 (2004).

\bibitem{Paredes}  B. Paredes {\it et al.}, Nature {\bf 429}, 277 (2004).

\bibitem{Kagome} L. Santos \textit{et al.}, Phys. Rev. Lett. {\bf 93}, 030601 (2004).

\bibitem{triangular} S. Wessel and M. Troyer, cond-mat/0505298.

\bibitem{Cirac} B. Paredes, C. Tejedor, and J. I. Cirac, cond-mat/0306497.

\bibitem{BG} B.\ Damski {\it et al.}, Phys. Rev. Lett., {\bf 91}, 080403 (2003);
R. Roth and K. Burnett, J. Opt. B Quant. Semiclass. Opt. {\bf 5}, S50 (2003).

\bibitem{Anderson} P. W. Anderson, Phys. Rev. {\bf 109}, 1492 (1958).

\bibitem{Fisher} M. P. A. Fisher {\it et al.}, Phys. Rev. B {\bf 40}, 546 (1989).

\bibitem{Inguscio} J. E. Lye, \textit{et al.}, cond-mat/0412167.

\bibitem{Sanpera} A. Sanpera,  \textit{et al.}, Phys. Rev. Lett. {\bf 93}, 040401 (2004).

\bibitem{Feshbach} E. Tiesinga {\it et al.}, Phys. Rev. A {\bf 47}, 4114 (1993); S. Inouye {\it et al.}, 
Nature (London) {\bf 392}, 151 (1998).

\bibitem{Adrian} In free space, 
a renormalization group analysis shows that the
critical properties at the transition between superfluid and disordered state
are unaltered as compared to the case of random $\mu$. A. Kantian, Dipl. Thesis, Univ. Hannover (2004), 
and private communication.

\bibitem{Schmiedmayer} S. Wildermuth {\it et al.}, Nature 435, 440 (2005).

\bibitem{footnote} As observed in Fig.~\ref{fig:fig1}, in practice the distribution of $U_i$ is neither 
ideally symmetric nor uniform, but this does not affect our main conclusions and could be easily included 
in both the SCE and QMC calculations. Also some residual noise correlations could appear between nearest 
neighbors, but they are small ($20\%$ in Fig.~\ref{fig:fig1}) and can be reduced by increasing the lattice spacing.

\bibitem{Svistunov} N. Prokof'ev and B. Svistunov, Phys. Rev. Lett. {\bf 80}, 4355 (1998); 
S. Rapsch, U. Schollw\"ock, and W. Zwerger, Europhys. Lett. {\bf 46}, 559 (1999); 
 N. Prokof'ev and B. Svistunov, Phys. Rev. Lett. {\bf 92}, 015703 (2004).

\bibitem{Monien} J. K. Freericks and H. Monien, Phys. Rev. B {\bf 53}, 2691 (1996).

\bibitem{sse} A.W. Sandvik, Phys. Rev. B {\bf 59}, R14157 (1999).
\bibitem{directedloop1} O. F. Sylju{\aa}sen and A. W. Sandvik, Phys. Rev. E {\bf 66}, 046701 (2002).
\bibitem{directedloop2} F. Alet, S. Wessel, and M. Troyer, Phys. Rev. E {\bf 71}, 036706 (2005).

\bibitem{footnoteT} We found a value of $T=0.015 J$ to be sufficient for the considered system sizes.

\bibitem{rho_S} E.L. Pollock and D.M. Ceperley, Phys. Rev. B {\bf 36}, 8343 (1987).

\bibitem{scalettar} R. T. Scalettar, G. G.  Batrouni, and G. T. Zimanyi, Phys. Rev. Lett. {\bf 66}, 3144 (1991).

\bibitem{Rigol} M. Rigol, A. Muramatsu, G. G. Batrouni, and R. T. Scalettar
Phys. Rev. Lett. {\bf 91}, 130403 (2003).

\bibitem{Wessel} S. Wessel, F. Alet, M. Troyer, and G. G. Batrouni,
Phys. Rev. A {\bf 70}, 053615 (2004).


\end{thebibliography}
\end{document}